# Room temperature magneto-optics of ferromagnetic transition-metal doped ZnO thin films

JR Neal[1], AJ Behan[1], RM Ibrahim[1], H J Blythe[1], M Ziese[2], AM Fox[1] and GA Gehring[1*]

[1]*Department of Physics and Astronomy, The University of Sheffield, Sheffield, S3 7RH, UK.*
[2]*Division of Superconductivity and Magnetism, University of Leipzig, D-04103 Leipzig, Germany*

## Abstract

Magneto-optic studies of ZnO doped with transition metals Co, Mn, V and Ti indicate a significant magnetic circular dichroism (MCD) at the ZnO band edge at room temperature, together with an associated dispersive Faraday rotation. Similar spectra occur for each dopant which implies that the ferromagnetism is an intrinsic property of the bulk ZnO lattice. At 10 K additional paramagnetic contributions to the MCD are observed, but above about 150K, the magnitude of the MCD signal is dominated by the ferromagnetism and is almost temperature independent. The MCD at the ZnO band edge shows room temperature hysteretic behaviour.

The search for spintronic materials that combine semiconducting and ferromagnetic properties, dilute magnetic semiconductors (DMS), is currently the most topical field in magnetism. The compounds based on the wide-gap (3.4 eV) semiconductor ZnO are especially exciting because they exhibit ferromagnetism at room temperature [1-6], in contrast to the GaMnAs-based materials for which the highest reported Curie temperatures, $T_c$, are still well below 300 K.

Despite the growing body of evidence in favour of room temperature magnetic hysteresis in doped ZnO, the nature of the ferromagnetism is hotly contested. The original interest in ZnO was prompted by the prediction that the hole exchange mechanism found in GaMnAs would produce Curie temperatures above 300 K [7]. However, it is now known that the doping is normally $n$-type, for which the exchange is smaller [8]. Furthermore, the measured moments are often much smaller than the theoretical values, which suggests that the magnetism might be due to an impurity phase [9,10]. It is therefore highly important to carry out careful experimental studies that can elucidate the microscopic origin of the magnetism.

In this Letter we present a detailed study of the magneto-optical (M-O) properties of ferromagnetic ZnO, together with other experimental details. A measurement of the magnetic circular dichroism (MCD) at photon energy $E$ gives the difference in absorption for left and right circularly polarised light at that same energy. Hence it provides a clear indication of the extent to which the states involved in the transition at that particular energy are influenced by the magnetism. We concentrate here on the spectral region close to the band edge at ~3.4 eV, since this characterizes the intrinsic behaviour of the ZnO lattice. The results show that the ferromagnetism at 300K is intimately connected with the band electrons of ZnO and that the carriers are polarised. Furthermore, since ZnO is transparent in the blue/UV, the large Faraday rotations that we observe around 3 eV are potentially useful for applications in M-O processing.

The $Zn_{1-x}M_xO$ samples studied here (M = transition metal) were grown as thin films by pulsed laser deposition (PLD) on sapphire (0001) substrates. Co, Mn, V and Ti were used as the dopants with concentrations up to 5%. An undoped ZnO film was also grown for reference. SQUID hysteresis loops of the doped films all show room temperature ferromagnetism, as has been observed previously [1,2,4]. Films with thicknesses in the range 200-500 nm were selected to permit partial transmission for light above the band gap. We agree with previous workers [2,6] that the highest magnetic moments are obtained in samples that are grown with an oxygen deficiency at low $O_2$ pressures, thus resulting in Zn interstitials or oxygen vacancies. However, different growth conditions are required to optimise the optical properties, and all the measurements presented here were made on films grown at 10 mTorr of $O_2$. At this pressure the sharpest spectral features were obtained [4], but the moment is smaller than those grown at lower $O_2$ pressure. The experimental values of the moment per transition metal ion (and the theoretical maxima) are Co: $0.45\mu_B$ ($3\mu_B$), Mn: $0.1\mu_B$ ($5\mu_B$), Ti: $0.06\mu_B$ ($2\mu_B$), V: $0.05\mu_B$ ($3\mu_B$).

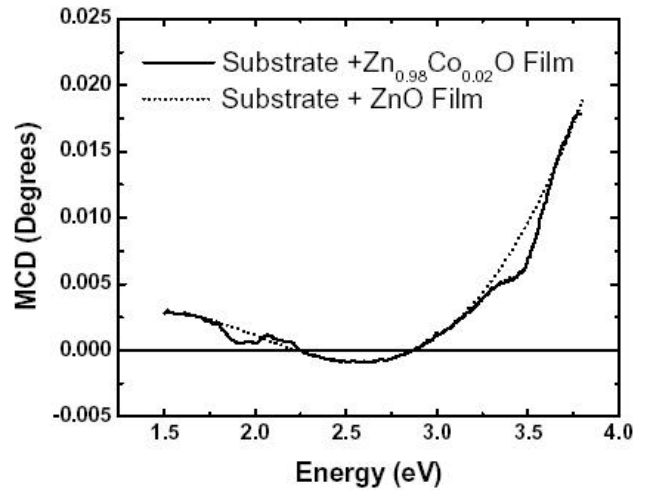

FIG. 1. MCD of an undoped ZnO film grown on a *c*-cut sapphire substrate compared to that of a 200 nm film of $Zn_{0.98}Co_{0.02}O$ grown on a similar substrate. Both sets of data were taken at 10 K in a field of 0.45 T. The arrows indicate the *d-d* transitions at 2 eV and the feature at the ZnO band edge at 3.4 eV for the $Zn_{0.98}Co_{0.02}O$ film.

The M-O spectra were taken with a xenon lamp and monochromator with a photoelastic modulator that allows for simultaneous recording of the Faraday rotation and MCD as a function of frequency. The Mn doped film was 520nm thick which resulted in a noisier spectrum in the MCD and hence the hysteresis loop was obtained in reflection (polar Kerr geometry). For low temperature measurements the samples were mounted in a cold finger cryostat with a temperature range of 10 K to 300 K. An electromagnet provided a field of 0.45 T with the cryostat and 0.9 T without the cryostat. Figure 1 presents typical raw data for an undoped ZnO film grown on a *c*-cut sapphire substrate and a 200-nm thick $Zn_{0.98}Co_{0.02}O$ thin film sample grown on a similar substrate. The MCD of the pure ZnO sample was dominated by the signal from the sapphire substrate of thickness 0.5 mm. The substrate signal varied linearly with the magnetic field and rather weakly with temperature and originates primarily from the tail of the band edge at 7 eV. The contribution of the magnetic film alone, as shown in Figs 2 and 4 below, was obtained by subtracting off the pure ZnO/substrate signal.

Figure 2 shows the room temperature MCD spectra of ZnO doped with Co (2%), Mn (2%), Ti (2%) and V (5%), together with the absorption spectra for the Mn, Co and V-doped samples, and the Faraday spectrum for the Ti-doped film. The absorption spectrum of the Ti:ZnO was similar to that of V:ZnO. The absorption spectra all show an edge at the band gap, although the Mn spectrum is much steeper than the other three. The sign of the MCD signal at the band edge was the same for all dopants consistent with an electron mechanism in all cases[11]We shall argue below that this is related to the fact that only the Mn-doped sample showed semiconducting rather than metallic conductivity at 300K.

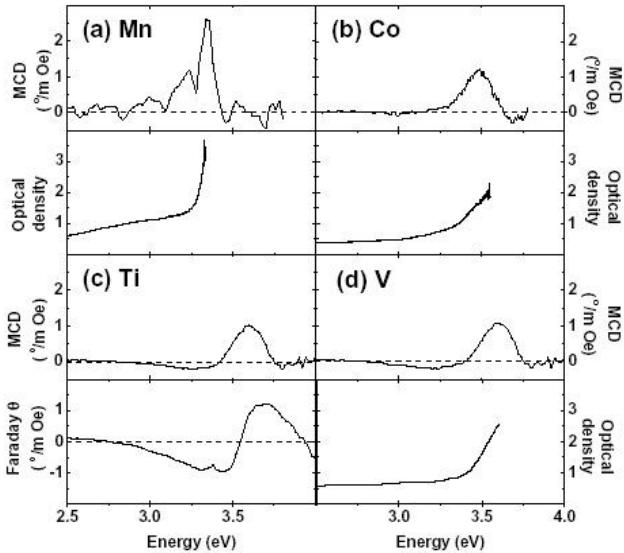

FIG. 2. Room temperature MCD measurements of ZnO doped with (a) Mn (2%), (b) Co (2%), (c) Ti (2%) and (d) V(5%). Optical absorption data for the Mn, Co, and V-doped films are also shown, together with Faraday rotation data for the Ti-doped film.

The key point to note in the data presented in Fig. 2 is that all the MCD spectra show a peak at the band edge. We can compare our data to that obtained for ferromagnetic GaMnAs, where a weak negative feature was observed between the band edge and the Fermi energy below the Curie temperature [12]. In our case we have a similar feature but of positive sign.

The ferromagnetic origin of the band edge MCD is confirmed by the observation of open MCD hysteresis loops for the Co, V and Ti-doped films and a Kerr rotation hysteresis loop for Mn. Figure 3 shows the open loop obtained for $Zn_{0.98}V_{0.02}O$ compared to the closed loop for the control pure ZnO film. These results constitute the first observation of open M-O loops for the ZnO band edge transition at room temperature. Previous hysteresis loops were either taken at low temperature [13,14] or showed no resonant signal at the band edge [15]. Coercive fields of ~90, ~150, ~100 and ~180 ±40Oe were measured for the Mn, Co, Ti and V doped samples respectively, and the values are consistent with those of ~100 Oe measured by SQUID magnetometry [1,2,3].

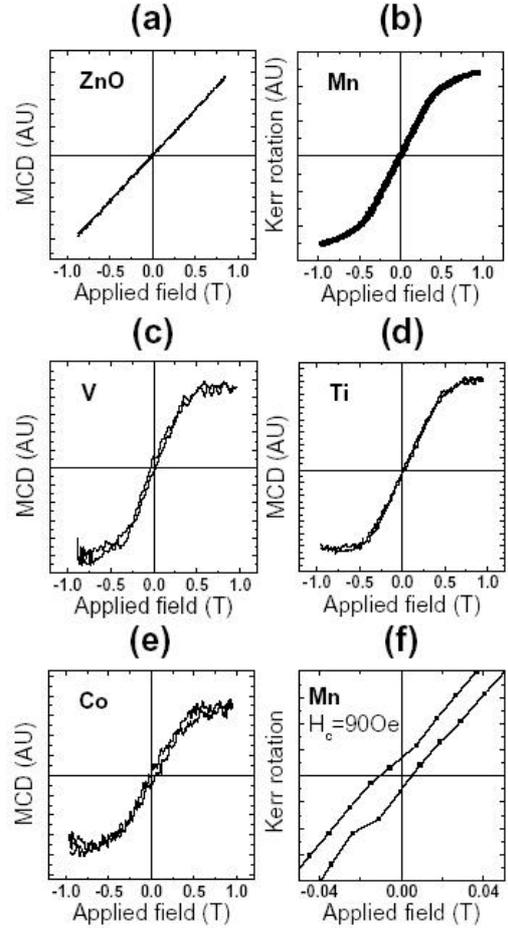

FIG. 3. MCD/Kerr hysteresis loops obtained at room temperature and an energy of 3.4 eV. (a) MCD versus field for the pure ZnO film. (b) Kerr rotation versus field for the same Mn doped film as Fig. 2a. (c)-(e) MCD versus field for the same V, Ti and Co doped films as Fig. 2b-d. (f) Expanded plot of (b) showing the coercive field. The linear variation of the pure ZnO film shown in (a) has been subtracted from the data and the graphs rescaled accordingly.

The origin of the MCD may be explored in more detail by studying its temperature dependence. Figure 4 presents MCD data for the Mn and Co-doped samples from 10 K to room temperature. Both samples show a negative excitonic feature at low temperatures that has been observed previously [13]. This signal weakens rapidly as the temperature increases both because of its paramagnetic nature and because the number of carriers increases, leading to exciton screening. The band edge signal at temperatures above 150 K is dominated by the positive feature shown in Fig. 2. The fact that the MCD does not show a dependence on $1/T$ above 100K is a further indication that it arises from the ferromagnetic component.

The spectrum of Mn-doped ZnO shows an additional feature at ~2.5 eV that is weakened at 50 K and has vanished by 100 K. This appears to correspond either to a minority phase of $Mn_3O_4$ which has a $T_C$ of 43 K or to $Mn_2O_3$, which has a transition at 83 K. At higher energy there are some very weak Mn *d-d* transitions between 2.55 eV and 3.2 eV [16]. At low temperatures the Co-doped material also shows an additional feature at 2 eV from crystal-field splittings of the $Co^{2+}$ ion [13].

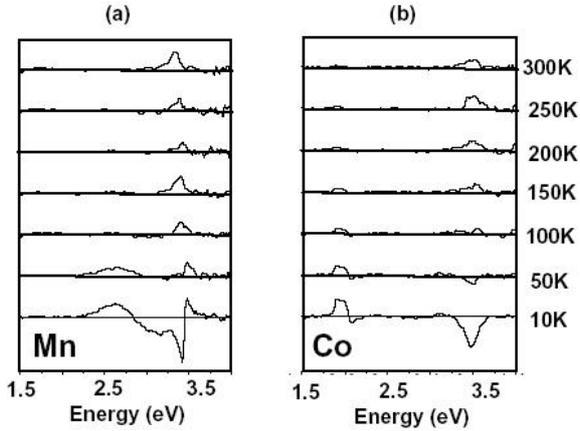

FIG. 4. Temperature dependence of the MCD of (a) $Zn_{0.98}Mn_{0.02}O$ and (b) $Zn_{0.98}Co_{0.02}O$.

Having presented the data, we can discuss the origin of the band edge features that are observed. We begin by considering the relationship between the MCD and Faraday data presented in Fig. 2. The MCD, $h$, and Faraday rotation, $q$, at angular frequency $w$ are proportional to the imaginary and the real parts of the off-diagonal component of the dielectric constants, $e_{xy}$, respectively and should satisfy a Kramers-Kronig relation. This is seen in all our samples, and demonstrated for the Ti doped film in Fig. 2(c). This verifies that the main contribution to the Faraday rotation originates from the same band edge transition that dominates the MCD.

Now let us consider the relationship between the optical data and the free carrier density. The absorption edge of heavily-doped semiconductors differs from a pure sample in two ways [17,18]. First, the Burnstein-Moss shift due to the extrinsic carriers has the effect of sharpening the absorption edge and raising it to higher energies, according to $\sqrt{E - E_g}\left[1 - f(E - E_F)\right]$, where $E_g$ is the band gap, $E_F$ is the Fermi energy and $f(E)$ is the Fermi–Dirac function. Second, the disorder causes band tails to form on both the valence and the conduction bands. In the case of doped ZnO, we are dealing with an $n$-type material, and $E_F$ will therefore lie in the conduction band. Hence the transitions that are observed to the vacant conduction band states start either from the valence band tail or from valence band states. In these circumstances one expects an absorption that varies linearly with energy for the valence-tail to conduction-band transitions and then a well-defined absorption plateau at higher energies above the onset for band-to-band transitions as sketched in Fig. 5a [17]. The spin-splitting of the bands shown in Fig. 5b causes a shift of the absorption plateau for $\sigma^+$ and $\sigma^-$ light as indicated in Fig. 5c. We then expect a peak of width $\sim \Delta$ at $E_g+E_F$ in the MCD as sketched in Fig 5d. The optical splitting $\Delta$ is related to the band splittings, $d_c$ and $d_v$, by $\Delta = d_c m_v / m_c - d_v$, where the subscripts refer to the conduction and valence bands respectively, and $m_i$ is the appropriate effective mass.

The schematic behaviour depicted in Fig. 5 is clearly observed in the experimental data for the Co, V and Ti samples given in Figs. 2 and 4. The carrier densities at room temperature in these samples were determined by Hall measurements to be: $n = 1.2 \times 10^{20}$ cm$^{-3}$ for Co, $n = 9.8 \times 10^{20}$ cm$^{-3}$ for Ti and $n = 1.1 \times 10^{21}$ cm$^{-3}$ for V. Thus in all three cases $E_F$ is considerably larger than $k_B T_c$, and an MCD peak at the band edge is observed in agreement with Fig. 5d. In principle, the width of the peak should give an estimate of $\Delta$. However, the transitions suffer from inhomogeneous broadening, as demonstrated by the broad Co crystal-field transitions at 2 eV. Hence the observed widths (~0.14 eV) only give an upper bound on the exchange splitting. We note that a band splitting of ~0.14 eV is actually much larger than would be required to produce room temperature ferromagnetism [6]. The data shown in Fig. 2 demonstrate that the main MCD peak is related to band-to-band rather than tail-to-band transitions because the peak comes close to the energy where the absorption is saturating rather than to the threshold energy for the tail-state absorption. This transition gives a weak positive MCD because the polarisation of the conduction band is positive and the spin-orbit interaction is weak.

It is apparent from Fig. 2 that ZnMnO differs significantly from the other three samples. This difference can be related to its much lower carrier density of $n = 1.0 \times 10^{17}$ cm$^{-3}$. All the samples were grown under the same conditions, and under these conditions a pure ZnO film should be expected to have a carrier density of ~$10^{17}$ cm$^{-3}$ [4]. The much larger carrier densities observed in Co, V and Ti samples are a consequence of the fact that the conduction band is hybridising with some of the $d$ electrons of the transition metal dopant. This should be the effect that drives the magnetism in these samples. The low carrier density in the Mn sample may be a consequence of the fact that the Mn $d$ electrons hybridise less strongly with the ZnO conduction band [2,5,11]. It is significant to note that ferromagnetism is still observed in the semiconducting Mn sample, with its relatively low density of mobile carriers, which implies that the ferromagnetism can be induced either by localised defect states or band states. The differences in the optical spectra (e.g. the steeper absorption edge) are consistent with the absence of significant band-tailing, and hence the increased importance of band-to-band transitions, as appropriate for the low carrier density.

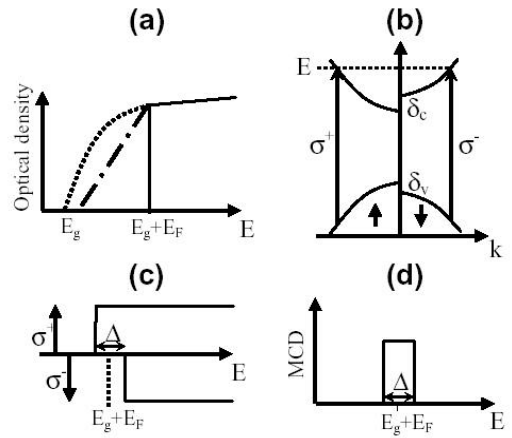

FIG. 5. (a) Sketch of the optical density in a non-magnetic, heavily-doped $n$-type semiconductor. Dotted line: undoped semiconductor. Dash-dotted line: absorption due to transitions from valence band tail states to the conduction band above the Fermi energy $E_F$. Solid line: band-to-band transitions above $E_F$. (b) Spin splitting of the conduction and valence bands, leading to different Fermi momenta and hence different energies for $\sigma^+$ and $\sigma^-$

transitions to conduction band states at $E_F$. (c) Band absorptions from (a) shifted to $E_g + E_F \pm \Delta/2$ for the opposite polarizations. (d) MCD expected from (c), showing a peak of width $\sim\Delta$ at $E_g + E_F$.

Finally, we note that the general agreement between the experimental data and the split-band model is good evidence that the ferromagnetism is not related to an impurity phase. Clustering of transition metal dopants would lead to strong antiferromagnetic exchange interactions across the oxygen bond, and this could explain the low magnetic moments observed in our samples. However, we would not then be able to explain the MCD peak at the band edge. Furthermore, we can argue that the ferromagnetism in $Zn_{1-x}M_xO$ is likely to involve only the fraction, $(1-x)^z$, of the transition metal, M, sites that do *not* have a magnetic neighbour, where $z$ is the number of nearest neighbours. With $z \sim 12$, values of $x < 6\%$ have at least half of the sites free to participate, which explains why low values of $x$ are required. Low temperature processing is also essential to avoid clustering [1].

In conclusion, the results that we have obtained show unambiguously that the ferromagnetism in doped ZnO causes band splitting in the bulk ZnO lattice. The particularly important features are the following:

(i) The room temperature MCD occurs at the ZnO band edge. A Faraday effect is observed above 3 eV in a region where the ZnO film is still transparent.

(ii) The *sign* of the MCD spectra indicates that the conduction band is hybridised with the magnetic ions and is spin split.

(iii) The fact that the behaviour is similar for different dopants indicates that the ZnO may be driven into a similar ferromagnetic state by different transition elements. (All the observed effects are certainly absent however for undoped ZnO).

(iv) The intensity of the MCD does not fall as the temperature is raised above 150 K. For the first time, open MCD hysteresis loops have been observed associated with the ZnO band edge at room temperature with coercive fields consistent with SQUID measurements.

(v) The Mn-doped sample differs from the other three in several ways: it is semiconducting, there is a narrower band of defect states showing up in the absorption, and the MCD effect itself is the largest by a factor of ~ 2.

(vi) We have shown that the splitting of the band states is independent of whether the magnetism occurs in films that are good conductors or rather poor conductors at room temperature.

We would like to thank Prof. Miroslav Kucera for advice and assistance in the early stages of the project, and we acknowledge support from the EPSRC via grants GR/S04352/01 and EP/D037581/1.

*Electronic address: G.A.Gehring@Sheffield.ac.uk